\documentclass[aps,pra,amssymb,bibnotes,twocolumn]{revtex4-2}
\usepackage{xcolor,color,graphicx,float,amsmath}

\begin{document}
\draft
\def\ds{\displaystyle}

\title{Subskin modes in a nonlinear non-Hermitian system}

\author{C. Yuce}
\affiliation{Department of Physics, Eskişehir Technical University, Eskişehir 26555, Turkiye}
\email{cyuce@eskisehir.edu.tr}

\date{\today}
\begin{abstract} 
Subskin modes are distinct from conventional skin modes as they localize not at the system's edge but rather below the edge. Unlike skin modes, where a substantial number of them can accumulate at the boundaries of a system due to the non-Hermitian skin effect, subskin modes are typically limited to one or a few and emerge only when the size of the system and the couplings are related in a very specific way. The nonlinear interaction can lift the restrictions on the couplings for the formation of these modes. The findings reveal the potential of subskin modes for practical applications and new research in non-Hermitian systems.
\end{abstract}

\maketitle

\section{Introduction}
Subsurface waves, also known as internal waves, represent a unique class of waves that propagate beneath surfaces. These waves have attracted a great deal of attention due to their significant impact on various natural systems. In geophysics, for example, seismic waves that travel through the Earth’s crust help us understand the planet's structure and the mechanisms underlying earthquakes \cite{internalw1}. Similarly, in oceanography, oceanic internal waves, which cause vertical movement of subsurface water layers while leaving the surface largely unaffected, play a vital role in marine ecosystems and global climate systems \cite{internalw2}. Subsurface waves can form due to various physical factors, such as density differences, buoyancy, dissipation, and nonlinear interactions. Understanding the underlying physics of these waves not only deepens our fundamental knowledge but also has the potential to open up exciting research opportunities in various fields of science. In one dimension, the stationary modes beneath the edges (subskin modes) are particularly fascinating. This paper investigates subskin modes in a one-dimensional non-Hermitian lattice and studies the effect of the Kerr-type nonlinear interaction on their formations.\\
The non-Hermitian skin effect typically implies localization at the edge and strong sensitivity of spectrum on the boundary conditions \cite{wangin,sw1}. We demonstrate that this effect can lead to subskin localization when the system has asymmetric long-range couplings, provided these couplings are strong and precisely controlled. Notably, subskin modes may appear even if the system exhibits the topological funneling effect, where an initial wave packet is directed toward one edge \cite{scifun,scifun2}. The restriction on the couplings for the emergence of these modes can be lifted by introducing Kerr-type nonlinearity into the system. Therefore, we say that the nonlinear non-Hermitian skin effect plays an important role on these waves. We note that recent researches have already extended the non-Hermitian skin effect into the nonlinear regime and showed intriguing results \cite{nonlin1,gsdye,nonlin2,nonlin4,nonlin5,nonlin6,nonlin7,nlekle1,nlekle2,nonlin9,sefbkj1,sefbkj2,sefbkj3,2eadkl,cemPRB2025,nlekle0,nonlin3,nonlinekleme}. For example, the emergence of the nonlinear skin effect was first explored in \cite{nonlin1}. The dynamical nonlinear higher-order skin effect \cite{nonlin2}, and nonlinear perturbations in systems with higher-order exceptional points \cite{nonlin4, nonlin5} enrich the understanding of this effect. A family of insensitive edge solitons in non-Hermitian topological lattices, which remain stable against perturbations is obtained \cite{nonlin6}. Recently, fixed points have been used as powerful tools for constructing nonlinear skin modes, which exhibit fascinating characteristics, such as degeneracy and power-energy discontinuities \cite{cemPRB2025}. Furthermore, it has been shown that the presence of an impurity can give rise to unique localized phenomena, such as discrete dark and anti-dark solitons as well as other localized modes that are neither skin modes nor scale-free localized modes. On the experimental side, the realization of the nonlinear non-Hermitian skin effect and skin solitons in an optical system with Kerr nonlinearity demonstrate skin effect in nonlinear domain \cite{nlekle0}.  Further experimental studies provided evidence of anomalous single-mode lasing driven by the combined effects of nonlinearity and non-Hermiticity \cite{nonlin3}. These developments collectively help us understand nonlinear non-Hermitian skin effect. \\
Subskin modes are elusive in the linear system as they are typically limited to one or a few and the couplings are required to have a size-dependent specific relation. On the other hand, they appear abundantly in the presence of the nonlinear interaction. The present study seeks to investigate these unique properties in a one-dimensional nonlinear and nonreciprocal lattice. \\
\section{Subskin modes}

The non-Hermitian skin effect leads to subskin localization beneath the edge under certain conditions. In a linear non-Hermitian lattice with asymmetric long-range couplings, a small number of modes may exhibit this localization. However, they appear more commonly in non-linear systems. To study subskin modes localized specifically below the left edge, we consider a one-dimensional non-Hermitian optical lattice with asymmetric long-range coupling in the backward direction and Kerr-type nonlinearity. The dynamical system is governed by the following equation for the complex field amplitude ${\Psi_n (z) }$ (${n = 1, 2, \ldots, N}$).
\begin{equation}\label{iqwaq300}
\sum_{m=1}^{s} J_m~ \Psi_{n+m} +J_{-1} ~\Psi_{n-1}  +g~|\Psi_n|^2\Psi_n  =i \frac{d\Psi_n}{dz} 
\end{equation}
where the propagation distance $z$ plays the role of time, $g>0$ is the nonlinear interaction strength, and $J_{-1},~J_1,~\ldots,~J_s$ are real-valued positive coupling constants with $J_s$ representing the longest-range coupling. By assuming a solution ${\Psi_n(z)= e^{-iEz} \psi_n}$, where $E$ is referred to energy for convenience, we obtain
\begin{equation}\label{iqwaq3f}
\sum_{m=1}^{s} J_m~ \psi_{n+m} +J_{-1} ~\psi_{n-1}  +g~|\psi_n|^2\psi_n  =E ~\psi_n 
\end{equation}
The system is subject to the open boundary conditions (OBC), where ${\psi_{0}=\psi_{N+1}=\psi_{N+2}=\ldots=\psi_{N+s}=0 }$, or to the semi-infinite boundary conditions (SIBC) when ${N\rightarrow\infty}$, where ${ \psi_{0}= \psi_{\infty}=0  }$. \\
Subskin modes are characterized by their confinement not at the edge but rather below the edge. Therefore, these modes require an additional condition beyond the boundary conditions (${\psi_1 = 0}$). In the linear system ${g=0}$, this extra condition, together with the OBC, imposes a  particular constraint on the couplings for the emergence of these modes. Fortunately, introducing nonlinearity removes this restriction, allowing them to emerge without specific coupling constraints. We note that deeper subskin modes (denoted by ${\psi_n^{(d)}}$) also exist, characterized by their localization further within the bulk such that ${\psi_1=\psi_2=\ldots=\psi_{d}=0}$, where ${d<s}$ is the depth from the left edge. Adding these extra conditions leads to more constraints on the couplings that cannot be lifted even by the nonlinearity. In other words, deeper nonlinear subskin modes require fine-tuning of the couplings to appear. Note that both skin and subskin modes arise from the non-Hermitian skin effect, but subskin modes can be localized far from the edge when $d$ is large.\\
We first revisit the non-Hermitian skin effect in the linear system ($g=0$), which occurs due to asymmetric couplings \cite{fannhse}. The emergence of this effect can be rigorously predicted by examining the spectrum under periodic boundary conditions,
${E_{PBC}=\sum_{m=1}^{s} J_m~e^{im k}+J_{-1} ~e^{-ik}}$, which becomes dense in the limit of ${N\rightarrow \infty}$, ${-\pi{\leq}k<\pi}$. This spectrum does not typically form conventional energy bands but instead exhibits loop-like structures in the complex energy plane as $k$ varies from ${-\pi}$ to ${\pi}$. Each loop is associated with a spectral winding number, given by ${	\nu = \int_{-\pi}^{ \pi} \frac{dk}{2\pi{i}} \frac{d }{dk} ln [ E_{PBC}-E_0 ]}$ for a complex reference energy $E_0$ inside the loop \cite{sw2,sw3,bipolar2,offf}. It is a topological invariant for the system, and a nonzero integer of $\ds{\nu}$ indicates the existence of the non-Hermitian skin effect in the system (see the insets of Fig. 1 for illustrations, where (a) shows two nested loops with ${\nu=1}$ for energies between the loops and ${\nu=2}$ inside the inner loop, and (b) shows more loops at $s=6$). Introducing boundaries requires a non-Bloch band theory to correctly describe the transition from the closed system to one with open boundaries. Remarkably, for the non-Hermitian system with spectral topology, the SIBC spectrum is exactly the regions with nonzero winding numbers (the interior of these spectral loops in the complex energy plane), while the OBC spectrum is a subset of the SIBC spectrum \cite{sw2,sw3}. Note that left-localized SIBC modes appear at energy $E$ inside the loop with a positive integer $\nu$ in the complex energy plane. \\
The nonlinear non-Hermitian skin effect remains poorly understood, and an analogous topological winding number is absent in the presence of nonlinear interactions. Notably, nonlinear systems display distinctive phenomena, such as fixed points and chaos. In the context of discrete nonlinear eigenvalue equations, a fixed point (also known as an equilibrium point) refers to a point in the system's state space at which the system remains indefinitely if it starts from this point. A fixed point is stable if the system returns to it after a slight perturbation. It is unstable if small disturbances will push the system away from the fixed point. Fixed points help determine the qualitative behavior of a nonlinear system. One important observation is that the zero fixed point can be used to construct nonlinear skin modes \cite{cemPRB2025}. In our system, we determine the fixed points by substituting ${\psi_{n}=a_0~ e^{i\theta}}$ into Eq.(\ref{iqwaq3f}), which yields ${ga_0^3=(E-E_c) a_0}$, where ${E_c=J_{-1}+\sum_{m=1}^s J_m}$ and $\theta$ is an arbitrary real parameter. Therefore, the system can have either a zero fixed point (${a_0=0}$), or nonzero fixed points (${ a_0^2=\frac{E-E_c}{g}}$ with ${E>E_c}$). Remarkably, the stability of the zero fixed point is influenced only by the system’s non-Hermiticity, while the stability of the nonzero fixed point is affected by both non-Hermiticity and nonlinearity. This can be seen from the linear stability analysis around the fixed points: ${\psi_n \rightarrow  a_0~ e^{i\theta}+ \epsilon~ \phi_{n}}$ with $|\epsilon| << 1$ and ${\phi_{\infty}=0}$. When ${a_0=0}$, this yields an equation that is independent of the nonlinear parameter: ${\sum_{m=1}^{s} J_m~ \phi_{j+m}+J_{-1}\phi_{j-1}=E~\phi_{j}}$, which satisfies the condition ${\phi_{\infty}=0}$ when $E$ is inside the spectral loops defined by $E_{PBC}$ in the complex energy plane ($\nu>0$). In contrast, when ${a_0\neq 0}$, both the couplings and the nonlinear parameter $g$ appear in the linearized equation. In this case, the nonzero fixed point loses its stability for large values of $E$, and bifurcations (period-doubling) occur until the system enters into a chaotic regime \cite{cemPRB2025}.

\subsection{Linear subskin modes}
The non-Hermitian skin effect may cause certain modes to localize beneath the edge. To gain insight, we provide an exact analytical solution under SIBC for a toy model with ${s=3}$ and ${J_2=0}$. Substituting ${\psi_n=\beta^n}$ into Eq. (\ref{iqwaq3f}) yields ${J_{3}~\beta^4+J_1~\beta^2+J_{-1}=E~\beta} $. The four roots of this polynomial equation are cumbersome to write here. At ${E=0}$, they simplify to $\beta_{\mp}$ and $-\beta_{\mp}$, where ${\beta_{\mp}=\sqrt{\frac{-J_1\mp~ \sqrt{J_1^2-4J_3~J_{-1}}}{2J_3}}}$. Therefore, when ${ \beta_{-}\neq \beta_{+} \neq 0}$, the zero-energy mode is given by a linear combination: ${\psi_n=(c_1+c_2(-1)^n)~\beta_+^n+(c_3+c_4(-1)^n)~\beta_-^n}$, where $c_{1,2,3,4}$ are constant coefficients with ${c_1+c_2+c_3+c_4=0}$ ($\psi_0=0$). Suppose that $J_3$ is large enough to ensure ${|\beta_{\mp}|<1}$ so that ${\psi_{\infty}=0}$ is satisfied. Therefore,  two of the constant coefficients in the solution remain as free parameters and can be chosen arbitrarily. In this way, we obtain two zero-energy subskin modes that are localized not at the left edge but below the edge with different depths
\begin{eqnarray}\label{2qdgafce}
\psi_n^{(1)} &=& c_1 (1+(-1)^n)~(\beta_+^n-\beta_-^n) \nonumber\\
\psi_n^{(2)} &=&  c_1 (1-(-1)^n)~(\beta_+^{n-1}-\beta_-^{n-1})  
\end{eqnarray}
where the constants $c_1$ are the corresponding normalization constants, and ${\psi_n^{(1)}}$ and ${\psi_n^{(2)}}$ represent the field amplitudes of the subskin mode localized just below the edge (${\psi_1=0}$) and at a deeper level (${\psi_1=\psi_2=0}$), respectively. For this example, the depth cannot be increased further. Note that the non-zero energy subskin modes can be constructed similarly by obtaining the four roots of the above polynomial equation for $\beta$ at a given value of $E$.\\
Introducing the right boundary into the system (${\psi_{N+1}=\psi_{N+2}=\psi_{N+3}=0}$) typically converts the subskin modes into skin modes. Fortunately, either solution in Eq. (\ref{2qdgafce}) can remain a valid solution under OBC, provided a specific restriction on the couplings is satisfied for a given lattice size (we refer to this as a rare solution, as it exists only under a specific condition). Assuming ${\beta_+^{\alpha}=\beta_-^{\alpha}}$ to satisfy OBC, where ${\alpha=N+2}$ (N:even) and ${\alpha=N}$ (N:odd) yields a restriction: ${J_1=0}$. In this case, ${\psi_n^{(1)}}$ and ${\psi_n^{(2)}}$ become eigenstates under OBC when ${N=4j+2}$ and ${N=4j+3}$, respectively (${j=1,2,...}$). Note that the OBC system still exhibits degeneracy at ${E=0}$. The other zero-energy mode is a skin mode. It is given by ${\psi_{n}=c_1~(1-(-1)^n)~(\beta_+^n- (\frac{\beta_+}{\beta_-})^{N+2} ~\beta_-^n) }$, where ${J=4j+1}$. We conclude that while subskin modes under SIBC can emerge over a wide range of coupling parameters, the open system exhibits a rare subskin solution, which occurs only when there is a restriction on the couplings. \\
The analytical solution obtained above demonstrates the existence of the SIBC and OBC subskin modes. For the generic model described by Eq.(\ref{iqwaq3f}), SIBC subskin modes appear at energy $E$ if there are at least three distinct nonzero roots of the polynomial equation ${ \sum_{m=1}^{s}J_m~\beta^{j+1} + J_{-1} = E~\beta}$, such that the absolute values of these roots are less than $1$. A subset of the SIBC subskin modes becomes OBC subskin modes. They are elusive because their existence depends on system size and requires a specific relationship between the couplings. Fig. 1 show the densities of some SIBC skin and subskin modes up to $n=20$ at $E=0.5$. For the system in (a), we numerically find ${\beta_1\approx-0.901}$, ${\beta_2=\beta_3^{\star}\approx(0.200 - 0.266~ i)}$, and hence we obtain ${\psi_n= c_1 \beta_1^n +c_2 \beta_2^n+c_3 \beta_3^n}$, where ${c_1+c_2+c_3=0}$ (${\psi_0=0}$) and ${\sum_n|\psi_n|^2=1}$. This solution satisfies ${\psi_{\infty}=0}$, and hence one of the constant coefficients ${c_{1,2,3}}$ remains a free parameter. By selecting it properly, we illustrate two skin modes (shown in black and blue) and a subskin mode (shown in red). The subskin mode has zero density at the left edge, but nonzero density below the edge with a decaying tail towards the right edge. In the inset, we plot the corresponding PBC spectrum, which consists of two nested loops. We numerically find that there are $3$ distinct values of $\beta$ with ${|\beta|<1}$ when the energy is inside the inner loop, and $2$ distinct values when the energy lies between the loops. Note that the winding number is $2$ for energies inside the inner loop, while it is $1$ for energies in the region between these two loops. As a next example, we illustrate deeper subskin modes. In Fig. 1(b), we set $s=6$ and $E=0.8$ and display the densities for a SIBC skin mode (in blue) and two SIBC subskin modes with different depths from the edge (in red and black). In this case, we carefully select the couplings ${J_{-1}}$ and $J_6$ so that the mode in red also becomes an OBC subskin mode for $N=10$, where its unnormalized form is given by $\ds{ \psi_n^{(2)}=\{  0,0,1, 2E,2E^2,E^3,0,-\frac{E^5}{2},-\frac{E^6}{2},-\frac{E^7 }{4}   \}   }$.  \\ 
\begin{figure}[t]
	\includegraphics[width=4.25cm]{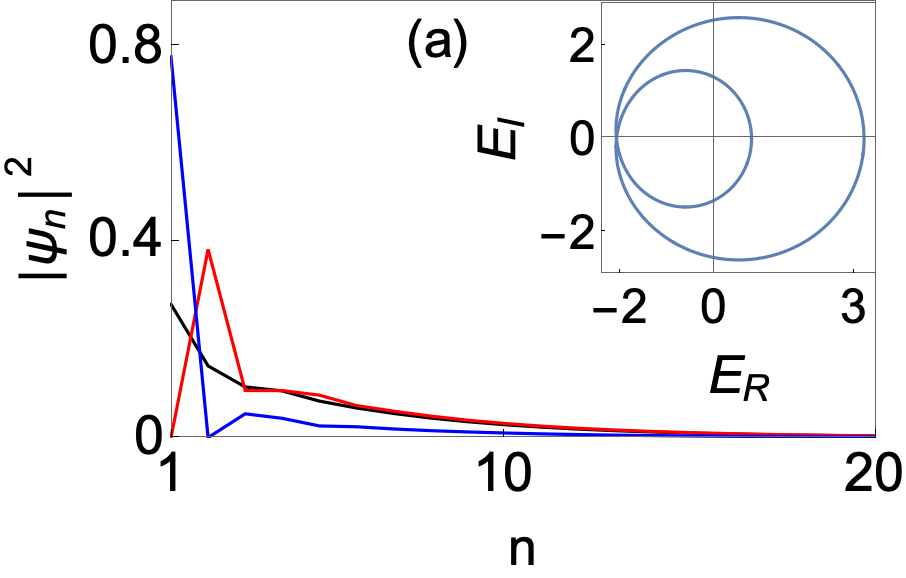}
    \includegraphics[width=4.25cm]{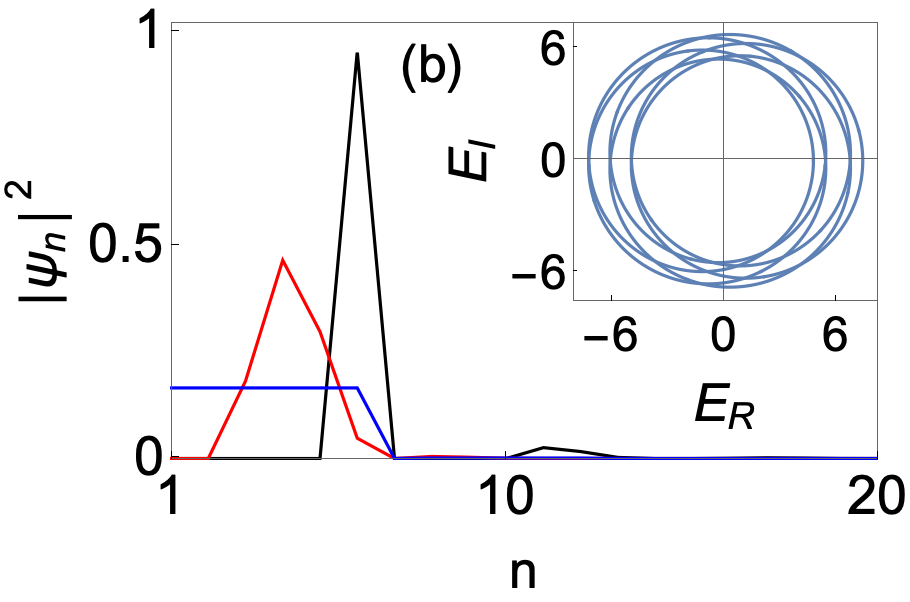}
	\caption{The densities of the SIBC skin and subskin modes for (a) ${J_2=10J_{-1}=2}$ at $E=0.5$, and (b) $\ds{J_6=\frac{1}{E^3~J_{-1} }=\frac{2}{E^5}}$ at $E=0.8$. The plots are displayed up to $n=20$ for clarity. The depth of the subskin mode is ${d=1}$ in (a), and ${d=2}$ and ${d=5}$ in (b). The skin mode in (b) has a large localization length with $\psi_{n\leq6}\approx0.407$. The insets show their PBC spectra. The common parameters are $J_1=1$ and $g=0$.}
\end{figure}
We have studied stationary subskin modes. To explore subskin waves propagating beneath the edge, we now consider a linear combination of subskin modes under SIBC: $\ds{ \Psi_n(z)= \sum_{d,E} c_{d,E}~e^{-iEz}~\psi_{n,E}^{(d)}  }$, where $c_{d,E}$ are constant coefficients, and the summation is over energy $E$ and depth $d$. Of particular importance is the consideration of real-valued energies, as such waves exhibit also power oscillation, which is a direct physical consequence of the non-orthogonality of the modes \cite{powoscu}. On the other hand, a real physical system has a finite length and OBC subskin modes are limited. Fortunately, we can approximately use the SIBC subskin modes if the lattice is large. This is because the influence of the right edge on the left-localized subskin waves remains nonsignificant up to a large propagation distance. As an illustration, consider a system with ${J_3=3}$, ${J_1=1}$, ${J_{-1}=0.1}$ and ${N=400}$. For this large size, we can practically use the SIBC modes ${ \psi_n^{(1)} }$ and ${ \psi_n^{(2)}}$, as the propagation is largely unaffected by the right edge up to a long distance. We select ${E=1}$ and ${E=-1}$ for ${ \psi_n^{(1)} }$ and ${ \psi_n^{(2)}}$, respectively, and form the initial wave as $\ds{\Psi_n(z=0)= \frac{\psi_{n,E=1}^{(1)}+ c ~\psi_{n,E=-1}^{(2)} }{\sqrt{1+c^2}}   }$, where $c$ is an arbitrary constant. Figure 2 plots the densities ${|\Psi_n(z)|^2}$ as a function of the propagation distance $z$ when ${c=0.5}$ (a) and ${c=1.5}$ (b). For clarity, the plots are displayed up to ${n=10}$. The maximum density is depicted in white, shifting to red as the density decreases, with black representing zero density. These waves exhibit power oscillations, which are more apparent around $n=5$ (the total power ${\sum_n |\Psi_n(z)|^2}$ oscillates between $1.00$ and ${1.15}$). These plots show that, despite dominant couplings toward the left edge, the wave packets propagate without reaching the left edge, i.e., ${\Psi_1(z)=0}$ for ${z<20}$. Note that at large propagation distance ${z>20}$, the wave packets are eventually deformed as a result of the finite length of the lattice. Therefore, they move to the left edge and remain there with growing power.  \\
Under OBC, subskin modes are typically limited to one or a few and require precise tuning of coupling parameters. As a result, realizing a subskin mode in a real physical setting is challenging, as small perturbations in the couplings cause the mode to rapidly move to the edge due to the topological funneling effect. Fortunately, these modes are abundant in the nonlinear systems. In the following, we demonstrate that the nonlinearity can induce the formation of subskin OBC modes with a depth of $d=1$, regardless of the lattice size or any specific relationships among the couplings.
\begin{figure}[t]
	\includegraphics[width=4.25cm]{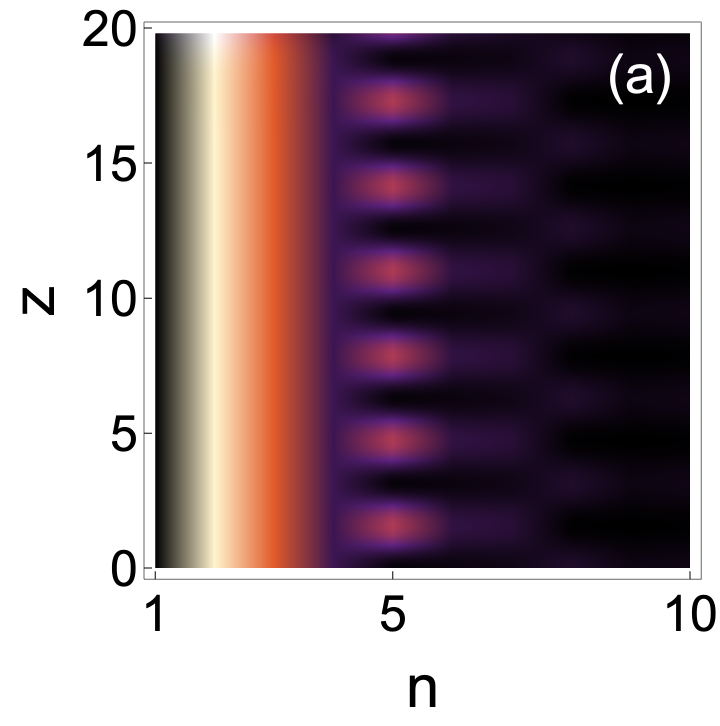}
    \includegraphics[width=4.25cm]{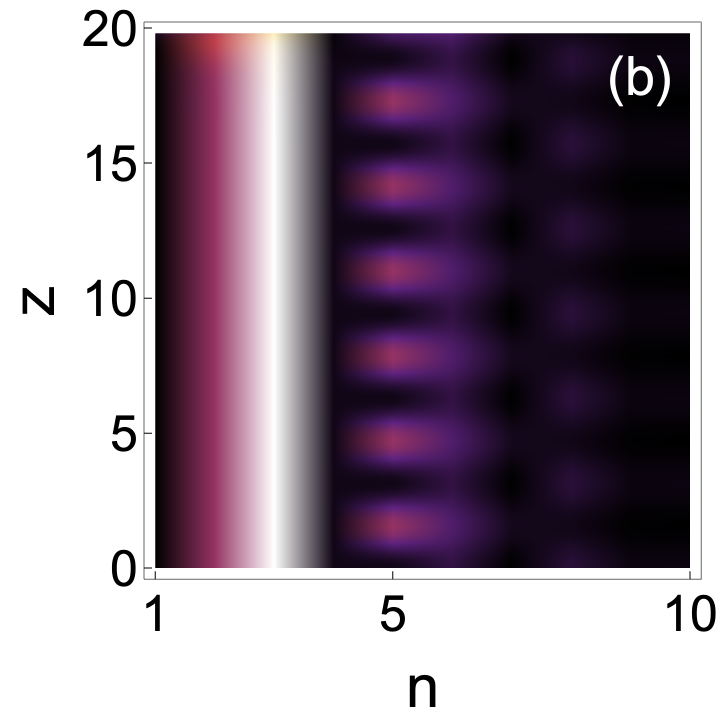}
	\caption{The density plots, ${|\Psi_n(z)|^2}$, illustrate the propagation of the initial wave function $\ds{\Psi_n(z=0)= \frac{\psi_n^{(1)} (E=1)+ c ~\psi_n^{(2)} (E=-1)}{\sqrt{1+c^2}}   }$ at (a) ${c=0.5}$ and (b) ${c=1.5}$. For clarity, the plots are displayed up to ${n=10}$. Both cases exhibit power oscillations, since the subskin modes are not orthogonal to each other. The parameters are ${J_{3}=3}$, ${J_{1}=1}$, and ${J_{-1}=0.1}$, $g=0$ and ${N=400}$.}
\end{figure}
\subsection{Nonlinearity}
The nonlinear interaction may induce stable subskin modes under OBC without requiring precise tuning of the couplings when ${d = 1}$, whereas deeper subskin modes typically emerge only under specific coupling relations. To clarify this issue, we first present exact analytical solutions for some small-sized lattices, then proceed with numerical computations for larger lattices. In the following, for the sake of simplicity, we consider real-valued field amplitudes.\\
For an analytical perspective, we set ${s=2}$ and ${N=4}$. The exact subskin solution of Eq.(\ref{iqwaq3f}) is given by ${
\psi_n^{(1)}=\{0, \psi_2, -\frac{J_1}{J_2}\psi_2,\frac{  J_1^2 + J_2 (E -   g  \psi_2 ^2) }{J_2^2} \psi_2\}} $, where ${J_2}$ is assumed to be much larger than ${J_1}$ and ${|E -   g  \psi_2 ^2|}$. The open boundary conditions at the right edge, ${\psi_5=\psi_6=0}$, impose the following two constraints: ${J_1^3 + J_{-1}J_2^2  + 2 J_1J_2 E - g \psi_2^2 (J_1J_2+   \frac{J_1^3}{J_2}) =0}$ and $ {J_1^4  + 3  J_1^2 J_2 E+J_2^2( 2 J_{-1} J_1 +  E^2)}-g \psi_2^2 ( J_1^2 J_2+ \frac{J_1^6}{J_2^3} + J_2^2 E+ 3 \frac{J_1^4}{J_2^2} E^{\prime} +  {E^{\prime}}^3 + \frac{J_1^2}{J_2}  (J_1^2 + 3 {E^{\prime}}^2))=0$ with $E^{\prime}= E - g \psi_2^2  $. From these constraints, it follows that ${E=\frac{(1-\sqrt{5})J_1^2}{2 J_2} }$ and ${J_{-1}= \frac{(\sqrt{5}-2)J_1^3}{J_2^{2}}}$ when ${g = 0}$, which restricts the values of one of the couplings for the appearance of that mode. However, the nonlinear interaction lifts this restriction on the couplings, allowing the subskin mode ${\psi_n^{(1)}}$ to appear over a wide range of couplings. This can be demonstrated by numerically solving the above constraints to determine the corresponding values of $E$ and $\psi_2$. For example, for ${J_1=10J_{-1}=J_2/5=1}$, the numerical solutions yield ${E=-0.67}$, ${\psi_2=0.79}$ when ${g=-1}$, and ${E=0.17}$, ${\psi_2=1.00}$ at ${g=1}$.\\
As a further example, we set ${s=3}$ and ${N=6}$ with ${J_2=0}$ to demonstrate the interdependence among the couplings necessary for the emergence of deeper nonlinear subskin modes. There are two subskin modes with distinct depths, ${\psi_n^{(2)} }$ and ${\psi_n^{(1)} }$. We start with the former one. Its field amplitude is given by ${\psi_n^{(2)}=\{
0, 0, \psi_3, 0, -\frac{J_1 \psi_3}{J_3}, \frac{ (J_1^2-J_3^2)g\psi_3^3}{2J_3^2} \}}$. The conditions at the right edges, ${\psi_7=\psi_8=\psi_9=0}$, yield three constraints: ${J_1^2 =J_{-1} J_3}$, ${E=\frac{  J_3^2+J_1^2}{2J_3^2}  g \psi_3^2}$ and $ J_1^3 J_3+ g \psi_3^2 (E - g \psi_3^2)^3 - J_3^2 (2 J_{-1}J_1  + E (E - g \psi_3^2))=0$. The first constraint restricts the possible values of one of the couplings, demonstrating that nonlinear subskin modes may not emerge without a coupling restriction when ${s>2}$. The last two constraints determine the values of $E$ and $\psi_3$. Note that these three constraints have no solution when ${g=0}$, which means that that mode is absent in the linear system. Next, we write the field amplitude of the other subskin mode, which is localized just below the edge. It is given by ${\psi_n^{(1)}=\{ 0, \psi_2, 0, -\frac{J_1 \psi_2}{J_3}, \frac{ (J_1^2-J_3^2) g\psi_2^3}{2J_3^3}, \frac{  (J_1^2-J_{-1} J_3) \psi_2}{J_3^2}\}}$, where ${J_3}$ is assumed to be much larger than $J_{-1}$ and $J_{1}$. The open boundary conditions at the right edges yield three constraints. They are given by ${ 2E= g \psi_2^2 (1 +  \frac{J_1^2}{J_3^2})}$, $ J_1^3 J_3+ g \psi_2^2 (E - g \psi_2^2)^3 - J_3^2 (2 J_{-1}J_1  + E (E - g \psi_2^2))=0$ and $EJ_3^4 (3 J_1^2 - 2 J_{-1}J_3)  - g \psi_2^2 ( J_1^2J_3^4 + J_1^6 -J_{-1} J_3^5  - 3 J_{-1} J_1^4  J_3 - J_{-1}^3 J_3^3  + J_3^2 (J_1^4 + 3J_{-1}^2 J_1^2 ))=0 $, which can be numerically solved to find $\psi_2$ and $E$. It follows that ${J_1^2 =2J_{-1} J_3}$ when ${g=0}$, but there is no such restriction when ${g\neq0}$. This shows that the nonlinear mode $\psi_n^{(1)}$ is available in this system without coupling restrictions, unlike the deeper mode $\psi_n^{(2)}$. \\
Obtaining analytical solutions for larger lattice sizes is challenging; therefore, we employ numerical analysis using {\it{shooting method}} \cite{cemPRB2025}. This method converts the boundary value problem into an initial value problem by guessing the field amplitudes near the left edge. In other words, to obtain ${\psi_n^{(d)} }$, we start with arbitrary initial values ${{\psi_{d+1},\psi_{d+2},\ldots,\psi_{s}}}$ and use them in the numerical iteration of Eq. (\ref{iqwaq3f}) to compute the subsequent terms up to ${\psi_{N+s} }$. This is repeated until the solution satisfies the boundary conditions at the right edge, ${\psi_{N+1}=\psi_{N+2}=\ldots=\psi_{N+s}=0 }$. For $s=2$, an alternative approach may be used within this method. Instead of an accurate initial estimate of ${\psi_2}$, one can plot ${\psi_{N+1}}$ and ${\psi_{N+2}}$ as functions of ${\psi_2}$ at fixed $E$. We then graphically identify particular values of ${\psi_2}$ for which ${\psi_{N+1}=\psi_{N+2}=0}$. Here, we consider real values of ${\psi_2}$ since replacing ${\psi_2e^{i\theta}}$ leads to ${\psi_n e^{i\theta}}$, where $\theta$ is an arbitrary real-valued constant. Extending this approach to systems with ${s>2}$ is challenging, as we may also need to guess the correct numerical values of the dependent couplings. Recall that deeper subskin modes appear when the couplings have a size-dependent specific relation.\\
We apply our approach to the system with ${N=60}$, ${J_2=2.3}$, ${J_1=1}$ and ${J_{-1}=0.1}$. We numerically see that no subskin mode exists when ${g=0}$ since linear OBC subskin modes only appear under a specific coupling relation that is not met in this case. However, in the nonlinear case with ${g=1}$, one can obtain a family of subskin modes by varying $\ds{E}$. For example, at ${E=0}$, we plot ${\psi_{N+1}}$ and ${\psi_{N+2}}$ as functions of $\psi_2$ in the inset of Fig.3(a) and find the values of $\psi_2$ where both are zero. We identify both the trivial value (${\psi_2 = 0}$) and the nontrivial values (such as ${\psi_2\approx 0.80765}$ and ${\psi_2\approx 1.33351}$). One can then numerically construct the modes ${\psi_n^{(1)}}$ by iterating Eq. (\ref{iqwaq3f}), starting with the nontrivial values of $\psi_2$ and using ${\psi_0=\psi_1=0}$ as initial conditions. Note that increasing the size of the lattice does not practically change these values of $\psi_2$; therefore, subskin modes can be obtained straightforward even for ${N>60}$. We can proceed by computing the time evolution of the modes obtained numerically. As an illustration, we compute the time evolution of the subskin mode at ${\psi_2\approx 0.80765}$. In Fig. 3(a), we plot the density ${|\Psi_n (z)|^2}$ as a function of the propagation distance by assuming ${\Psi_n (z=0)= \psi_n^{(1)}}$. The mode is localized beneath the edge, with a rapidly decaying tail towards the right edge. We see that it remains stationary up to ${z \approx 80}$. Note that enhancing the accuracy for the value of $\psi_2$ enables us to obtain stationary modes over a longer duration. Small fluctuations in numerical accuracy lead to deformation of the mode at later times. Once the deformation begins, the wave starts to expand towards both edges with increasing power. Note that the same initial wave packet in the absence of the nonlinear interaction would rapidly move to the left edge due to the topological funneling effect and remain localized at the left edge. \\
The illustration above shows that the shooting method provides us an easy way to obtain subskin modes for large lattice sizes when ${s=2}$. However, when ${s>2}$, implementing the shooting method becomes challenging due to the need for precise guesses of the couplings in addition to the initial values. Fortunately, we can approximately find the modes that remain nearly stationary over an extended propagation distance (quasi-stationary subskin modes). To do this, we relax the boundary conditions at the right edge so that they are not exactly zero but a tiny number, i.e. ${\psi_{N+1}\approx\psi_{N+2}\approx\ldots\approx\psi_{N+s}\approx0 }$. As an illustration, we consider ${s=4}$, and find a quasi-stationary subskin mode ${\psi_n^{(3)}}$. We plot the terms ${\psi_{N+1},\psi_{N+2},\psi_{N+3}  }$ and $\psi_{N+4}$ as functions of $\psi_4$ in the inset of Fig. 3(b), using different colors to distinguish them, at $g=1$. As can be seen, not all of them are zero at the same value of $\psi_4$. Fortunately, they are all less than ${10^{-20}}$ at ${\psi_4\approx 0.39}$. Therefore, the corresponding solution can be considered to be the one that survives sufficiently long. To check this idea, we compute its time evolution and plot the corresponding density in Fig.3(b). As seen, the wave packet propagates beneath the edge without being deformed up to ${z=80}$. It the moves to the left edge since the initial wave packet is quasi-stationary.\\
\begin{figure}[t]
	\includegraphics[width=4.25cm]{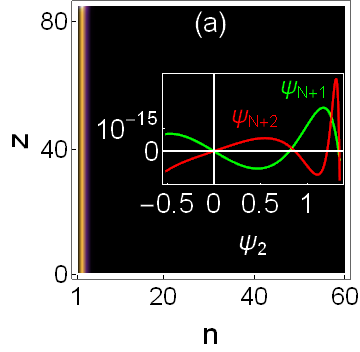}
    \includegraphics[width=4.25cm]{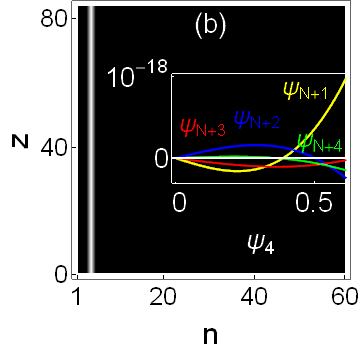}
	\caption{The densities $|\Psi_n(z)|^2$ as a function of time when the zero-energy subskin mode is initially excited, (a) ${\psi_1=0}$ and ${\psi_2\approx 0.80765}$, and (b) ${\psi_1=\psi_2=\psi_3=0}$ and $\psi_4=0.39$. The initial waves preserve its form up to $z\approx80$ before being deformed. The values of ${\psi_2}$ in (a) and ${\psi_4}$ in (b) are determined from the open boundary conditions at the right edge (see the insets). The parameters are (a) ${J_1=1}$ and ${J_2=2.3}$, and (b) ${J_4=4}$, ${J_{3}=J_{2}=0}$, ${J_1=0.1}$. The common parameters are ${J_{-1}=0.1}$, $g=1$ and $N=60$. }
\end{figure}
\section{Conclusion}
The non-Hermitian skin effect leads to the localization of eigenstates at or near the boundary of the system. Skin modes are boundary-localized states with nonzero amplitudes at the edge of the system. In contrast, subskin modes are localized states confined just beneath the edge. With sufficiently strong long-range couplings, deeper subskin modes can emerge, whose localization centers lie further within the bulk. In linear OBC systems with asymmetrical couplings, a large proportion of the modes can be skin modes. In contrast, subskin modes are typically limited to one or a few and emerge only when the size of the system and the couplings are related in a very specific way. As a result, small perturbations (violating this specific constraint) in a real physical setting cause them to move rapidly to the left edge due to the topological funneling effect. Note that a linear combination of subskin edge modes provides the most general form of subskin waves. They are characteristically different from the subsurface waves studied in the literature as they are mainly due to the non-Hermitian skin effect.\\
The nonlinearity can induce the formation of subskin modes under OBC without such a precise control of the coupling parameters. Specifically, subskin modes localized just below the edge (${d=1}$) may appear regardless of the specific values of the couplings. However, deeper nonlinear subskin modes exist under a specific relationship between the couplings. We discuss that the shooting method has been shown to be an easy and powerful method to find subskin modes with a depth of ${d=1}$.\\
This study was supported by Scientific and Technological Research Council of Turkey (TUBITAK) under the grant number 124F100. The authors thank to TUBITAK for their supports.

\end{document}